\documentclass[twocolumn,nofootinbib,floatfix]{revtex4-1}
\usepackage{graphicx}
\usepackage{xspace}
\usepackage[cp1251]{inputenc}
\usepackage[english,russian]{babel}

\usepackage{amsmath,amssymb}% fontenc, times, mathptmx}

\usepackage[dvips]{color}

\newcommand\black{\color{black}}

\usepackage{latexsym,graphicx,epsfig}
\def\lb{\linebreak[4]}
\newcommand{\be}{\begin{equation}}
\newcommand{\ee}{\end{equation}}
\newcommand{\bea}{\begin{eqnarray}}
\newcommand{\eea}{\end{eqnarray}}
\newcommand{\bes}{\begin{subequations}}
\newcommand{\ees}{\end{subequations}}
\newcommand{\bear}{\begin{equation}\begin{array}}
\newcommand{\eear}[1]{\end{array}\label{#1}\end{equation}}
\newcommand{\fr}[2]{\dfrac{{ #1}}{{ #2}}}
\newcommand{\pa}{\partial}
\newcommand{\la}{\langle}
\newcommand{\ra}{\rangle}
\newcommand{\fn}[1]{\footnote{{#1}}}
\newcommand{\bu}{$\bullet$\ }

\def\vep{{\varepsilon}}
\newcommand{\epe}{\mbox{$e^+e^-\,$}}
\newcommand{\ggam}{\mbox{$\gamma\gamma\,$}}

\def\cl{\centerline}

{\end{list}}
\newcounter{enumct}

\newcommand{\HB}{Higgs basis\xspace}
\newcommand{\LL}[1]{\Lambda_{#1}}
\newcommand{\ls}[1]{\lambda_{#1}}
\newcommand{\ff}[2]{\left(\!\phi_#1^\dagger\phi_#2\!\right)}
\newcommand{\fft}[2]{\left(\!\Phi_#1^\dagger\Phi_#2\!\right)}%new

\newcommand{\fftc}[2]{\left(\!\!\Phi_#1^\dagger\Phi_#2\!\!-\!\!\fr{v^2}{2}\!\!\right)}%new

\newcommand{\Ls}[1]{\Lambda_{#1}}

\begin{document}
\renewcommand{\tilde}{\widetilde}%[1]

\date{}
\title{
2HDM in terms of observable quantities}
\author{I.~F.~Ginzburg, K.~A.~Kanishev}
\affiliation{ Sobolev Institute of Mathematics, Novosibirsk, 630090, Russia; Novosibirsk State University, Novosibirsk, 630090, Russia}

% \maketitle

%%%%%%%%%%%%%%%%%%%%%%%%%%%%%%%%%%%%%%%%%%%%%%%%%%%%%%%%%%%%%%%%%%%%%

\begin{abstract}

{
 We found a minimal and a comprehensive set of directly measurable quantities defining the most general Two-Higgs-Doublet Model (2HDM), we call  these quantities {\it observables}. The parameters of potential of the model are expressed explicitly via these observables (plus non-physical parameters similar gauge parameters). The model with arbitrary values of these observables can in principle be realized (up to general enough limitations). Our results open the door for the study of   Higgs models in terms of  measurable quantities only. The experimental limitations can be implemented here directly, without complex, often model-dependent, analysis of  the   Lagrangian coefficients.

The principal opportunity to determine all parameters of 2HDM from the (future) data meets  strong practical limitation. In the best case it is the problem for very long time.}

Apart from this construction {\em per se}, we also obtain some  by-products.   { Among them -- a simple criterium for CP conservation in  the 2HDM},
a  new sum rules for Higgs  couplings, a clear possibility of the coexistence of relatively light Higgses with the strong interaction in the Higgs sector and {  a
simple expression for the  triple Higgs vertex $g(h_ah_ah_a)$, useful for  the analysis of future $hhh$   coupling  measurements.}

\end{abstract}

\maketitle

    %%%%%%%%%%%%%%%%%%%%%%%%%%%%%%%%%%%%%%%%%%%%%%%%%%%%
\section{Introduction} \label{ch.intro}
%%%%%%%%%%%%%%%%%%%%%%%%%%%%%%%%%%%%%%%%%%%%%%%%%%%%

The recent discovery of a Higgs-like particle with\lb $M\approx 125$~GeV at the LHC \cite{125Higgs} hints that the spontaneous electroweak symmetry breaking is most probably realized by the Higgs mechanism. The minimal realization of the Higgs mechanism introduces a single scalar isodoublet $\phi$ with the Higgs potential $V_H=-m^2(\phi^\dagger \phi)/2 +\lambda(\phi^\dagger \phi)^2/2$. This model is usually called ``the Standard Model'' (SM); it can be also referred as the ``one Higgs doublet model''.
The experimental results favor the realization of that minimal scenario \cite{125_2HDM} ({\it SM-like scenario { \cite{GKO}, or  SM alignment limit} \cite{Pilaf15}}). Nevertheless, many variants of extended Higgs models are not ruled out.

The Two Higgs Doublet Model (2HDM) presents the simplest extension of the standard Higgs mechanism. { This name unites  a  group of models in which} standard Higgs doublet is supplemented by an extra hypercharge-one doublet. It offers a number of phenomenological scenarios with different physical content realized in different regions of the model parameter space, such as a natural mechanism for spontaneous CP violation \cite{TDLee}. Some of its variants have a number of interesting cosmological consequences \cite{GIK09}.   The  Higgs sector of the MSSM is a particular case of 2HDM, etc.
{ Below we use the term 2HDM for the most general Two Higgs Doublet Model -- independently on possible CP violation, violation of $Z_2$ symmetry and with arbitrary Yukawa sector, etc.}

{ After EWSB the 2HDM contains 3 neutral Higgs bosons $h_a\equiv h_{1,2,3}$  and charged Higgs boson $H^\pm$ with masses $M_a$, $M_\pm$ respectively.}

In the SM, parameters of Higgs potential can be treated as measurable quantities, the mass of the Higgs boson $M_h$ and the Higgs self-coupling parameter\lb $\lambda=M_h^2/v^2$, where $v=246$~GeV is vacuum expectation value of Higgs field. Physical problems in this model can be equally discussed in terms of parameters of the potential or in terms of these observables.

2HDM contains two fields with identical quantum numbers. Therefore, its description in terms of  original fields or in terms of their linear superpositions   are equivalent;  this statement verbalizes the reparameterization (RPa) freedom of the model. A restricted version of this freedom, in which one only changes the phases of individual doublets, is called the rephasing (RPh) freedom. {
This freedom makes clear that the study in terms of Lagrangian
may be likened to discussion of electrodynamical effects in a certain gauge defined by  some  particular gauge-fixing conditions. The discussion of 2HDM in terms of only well measurable quantities seems preferable.}

{ The approach of many authors  is to express measurable quantities via parameters of Lagrangian (in some  RPa basis, often -- with some  simplifying assumptions) and to analyse physical phenomena via these RPa dependent parameters\fn{ In the analysis of data the parameters of Higgs Lagrangian are being obtained by a  solution of complex enough  set of  equations. The problem of  completeness, or predetermination of the observations that are necessary   for complete description of  the potential    is usually not discussed.} (see e.g. \cite{modscan}, \cite{softZ2-parameters}).  The  basis-independent approach of \cite{basindep}  is often free from  simplifying assumptions} but involves very bulky calculations.  The algebraic approach based on bilinears developed in \cite{bilinears}  allows one to deduce general properties of the models but has no advantages for analysis of phenomenology.

\bu {  Our approach is opposite. We have developed a method for finding  the minimal and a comprehensive set of directly measurable quantities defining the  2HDM and have built simple example of  such set. Further  we call  these quantities {\it observables} and call the chosen complete set   {\it The basic set of observables}. {
This basic set is subdivided naturally into two subsets, defined below.}
We have found simple explicit expressions for the parameters of potential of the model via these observables (and non-physical parameters, fixing RPa basis). (In this calculation  we continue  the earlier studies of ref.~\cite{CouplingCP}.)} { This approach allows to analyse physical phenomena only in terms of measurable quantities, without using of RPa dependent parameters of potential. Fortunately, the obtained description appeared to be simple enough.}

\bu The structure of the paper is the following. { In Sec.~\ref{ch.THDM} we reproduce well known facts in the useful for us form. We start with a brief review of the 2HDM and introduce useful notations. Next we turn to the Higgs basis, in which only one Higgs field have non-zero v.e.v. and which proves to be  very useful for our analysis. { In Sec.~\ref{ch.R} we consider quadratic (mass)
terms of the potential.} We start with the neutral components of the  original Higgs doublets and construct the physical Higgs fields. In the \HB, the components of the corresponding rotation matrix represent measurable physical quantities. { As a result, we express some of the parameters of the potential via the observables of the first subset --  Higgs boson masses and their couplings to $W^\pm$}. { The study of triple and quartic interactions of Higgs fields is necessary to describe the remaining parameters of potential. It is done in  Sec.~\ref{ch.self}, where we choose measurable vertices $g(H^+H^-h_a)$ and $g(H^+H^-H^+H^-)$ for the second subset of observables.} In Sec.~\ref{ch.Yuk}, we consider Yukawa couplings for each type of fermions.}
{ In the Sec.~\ref{chappl} we present some applications of obtained results and a number of useful by-products. Among them
}\\{ $\triangledown$ a simple criterium for CP conservation in  the 2HDM, written via measurable quantities only;\\
$\triangledown$  a  new sum rules for Higgs  couplings;}\\
{ $\triangledown$ a simple expression for the  triple Higgs vertex $g(h_ah_ah_a)$, useful for  the analysis of future $hhh$   coupling  measurements (in the separate paper);}\\
{ $\triangledown$  we have found the opportunity for co-existence of the relatively light Higgses and the strong interaction in the Higgs sector.}\\
 In Sec.~\ref{ch.disc} we discuss the results obtained.

%%%%%%%%%%%%%%%%%%%%%%%%%%%%%%%%%%%%%%%%%%%%%%%%%%%%
\section{Two Higgs Doublet Model} \label{ch.THDM}
%%%%%%%%%%%%%%%%%%%%%%%%%%%%%%%%%%%%%%%%%%%%%%%%%%%%
%\subsection{Generalities}\label{ch.THDMgen}
%%%%%%%%%%%%%%%%%%%%%%%%%%%%%%%%%%%%%%%%%%%%%%%%

The 2HDM describes a system of two scalar isospinor fields $\phi_1$, $\phi_2$
with hypercharge $Y=1$.  The most general form of the 2HDM potential is
\bear{rl}
V\! & \!=\!\fr{\ls1}{2}\ff11^2+\fr{\ls2}{2}\ff22^2+\ls3\ff11\ff22         \\[3mm]
  & +\ls4\ff12\ff21 + \fr{\ls5}{2}\ff12^2 + \fr{\ls5^*}{2}\ff21^2   \\[3mm]
  & +\left[\ls6\ff11\ff12+\ls7\ff22\ff12+\text{h.c.}\right]         \\[3mm]
 \! & \! -\fr{m_{11}^2}{2}\!\ff11\!-\!\fr{m_{22}^2}{2}\!\ff22\!-\!\!\left[\fr{m_{12}^2}{2}\ff12\!+\!\!\text{h.c.}\right]\!
\eear{thdmpot}
Its coefficients are restricted by the requirement that the potential be positive at large quasiclassical values of $\phi_i$
({\it positivity constraints}).

\bu  The model contains two  doublets of scalar fields with identical quantum numbers. Therefore, it can be described either in terms of the original fields $\phi_1$, $\phi_2$, which enter \eqref{thdmpot}, or in terms of fields $\phi'_1$, $\phi'_2$, which are obtained from $\phi_k$ by a global unitary {\it reparameterization} transformation ${\cal {\hat F}}$  of the form:
\bear{c}
\begin{pmatrix}\phi_1'\\ \phi_2'\end{pmatrix} =
\hat{\cal F}_{gen}(\theta, \tau,\rho)\begin{pmatrix}\phi_1\\ \phi_2\end{pmatrix}\,,\\[5mm]
\hat{\cal F}_{gen}=e^{-i\rho_0}\begin{pmatrix}
\cos\theta\,e^{i\rho/2}&\sin\theta\,e^{i(\tau-\rho/2)}\\
-\sin\theta\,e^{-i(\tau-\rho/2)}&\cos\theta\,e^{-i\rho/2}
\end{pmatrix}.
\eear{reparam}
This transformation induces a transformation of the parameters of the Lagrangian  $\lambda_i \to \lambda'_i$ in such a way that the new Lagrangian, written in fields $\phi'_i$, describes the same physical content.
We refer to these different  choices as different RPa bases.

Transformation \eqref{reparam} is parameterized by  angles
$\theta,\, \rho,\,\tau$ and $\rho_0$. The parameter $\rho_0$  describes an overall phase transformation of the fields,
and since it does not affect the parameters of the potential, we do not consider this degree of freedom.

In the potential \eqref{thdmpot}, parameters $\ls{1-4},\,m_{11}^2$ and $m_{22}^2$ are real while $\ls{5-7}, m_{12}^2$ are generally complex.
So, it takes 14 real quantities to fully define the scalar part of 2HDM.
Since the three remaining parameters of RPa transformation cannot influence description of physical phenomena, the actual number
of  physically relevant parameters of the potential is $14-3=11$.

\bu   Extrema of the potential satisfy the  stationarity equations $\left . \pa V/\pa \phi_i\right|_{\phi_1=\la\phi_1\ra, \phi_2=\la\phi_2\ra}=0$  ($i=1,\,2$).
The most general solution that describes the\lb $SU(2)\times U(1)_{Y} \to
U(1)_{EM}$ symmetry breaking can be expressed via two  positive numbers
$v_i$ and the relative phase factor $e^{i\xi}$ as:
\begin{eqnarray}
\langle\phi_1\rangle =\fr{1}{\sqrt{2}}\left(\begin{array}{c} 0\\ v_1\end{array}\right), \;\;
\langle\phi_2\rangle =\fr{1}{\sqrt{2}}\left(\begin{array}{c}0 \\ v_2 e^{i\xi}\end{array}\right)\,,\\[2mm]
v_1=v\cos\beta,\quad v_2=v\sin\beta,\quad v=\sqrt{v_1^2+v_2^2}\,.
\label{genvac}
\end{eqnarray}
The ground state of potential (the vacuum) is the extremum with the lowest energy, and its vacuum expectation value (v.e.v.) is
$v=246$~GeV.  $\phi_i$ are then decomposed into their v.e.v.'s and the quantized component fields:
\be
\phi_1 =\!\begin{pmatrix}\xi_1^+\\\dfrac{v_1+\zeta_1\!+\!i\xi_1}{\sqrt{2}}\end{pmatrix},\,
\phi_2 =\!\begin{pmatrix}\xi_2^+\\\dfrac{v_2+\zeta_2\!+\!i\xi_2}{\sqrt{2}}\end{pmatrix}e^{i\xi}\,.\label{videf}
\ee
Here, $G^\pm=\xi_1^\pm\cos\beta+\xi_2^\pm\sin\beta$ and  $
G^0=\xi_1\cos\beta+\xi_2\sin\beta$ are the massless Goldstone modes,
while\lb   $H^\pm=\xi_2^\pm\cos\beta-\xi_1^\pm\sin\beta$ and $\zeta_3=\xi_2\cos\beta-\xi_1\sin\beta$
describe  the charged Higgs boson and a neutral scalar $\zeta_3$ whose parity is opposite to that of $\zeta_{1,2}$.
Linear combinations of neutral fields
$\zeta_i$ form the set of observable neutral Higgs particles $h_1,\,h_2,\,h_3$.\\

%%%%%%%%%%%%%%%%%%%%%%%%%%%%%%%%%%%
%\subsection{ Relative couplings}\label{chrel}
%%%%%%%%%%%%%%%%%%%%%%%%%%%%%%%%%%

\bu {\bf  Relative couplings.}
{ Let us denote the coupling of each neutral Higgs boson to a fundamental particle $P$ by $g^P_a$ ($P=\{V(W,\,Z)$, $f=q(t,b,c,...),\,\ell(\tau,\mu, e)\}$) and these very couplings of the standard Higgs boson of SM as $g^P_{\rm SM}$. Below, we make use of the relative couplings, the ratios of these couplings:}
\be
\chi^P_{a}=g^P_a/g^P_{\rm SM}\,.\label{relcoupldef}
\ee

The model contains  an extra scalar-vector boson interactions, $H^\pm W^\mp h_a$ { and interactions $H^+H^-h_a$}.  For these we
introduce dimensionless relative couplings:
\bea
    &\chi_{a}^{H^+ W^-} = \dfrac{g(H^+ W^- h_a)}{M_W/v}\equiv \left(\chi_{a}^{H^- W^+}\right)^*\,;& \label{relcoupb}\\
&{\chi^\pm_a=g(H^+H^-h_a)/(2M_\pm^2/v)\,.}&\label{relcouplc}
\eea

{ The neutrals $h_a$  generally have no definite CP parity. Couplings $\chi^V_a$ and $\chi^{\pm }_a$ are real due to Hermiticity of Lagrangian, while other couplings  are generally complex.}

Relations among  relative couplings appearing in particular models are more stable under radiative corrections
than the relations among couplings them\-sel\-ves, since  possible large QCD corrections in nominator and denominator
of \eqref{relcoupldef} compensate each other.

{ We  omit the adjective "relative" further in the text.}

%%%%%%%%%%%%%%%%%%%%%%%%%%%%
%\subsection{Higgs basis}\label{ch.rpat}
%%%%%%%%%%%%%%%%%%%%%%%%%%%

\bu {\bf Higgs basis}.
Any RPa basis can be used for solving physical problem.
Some of them are more suitable than others when solving specific problems. In particular, when the system possesses an additional symmetry, the preferable RPa basis is the one in which this symmetry is  made obvious. Examples include the case when, in some RPa basis, the Yukawa sector has a form in which fermions of each type are coupled to only one field $\phi_1$ or $\phi_2$ (well known models I, II, X, Y); the case of softly broken $Z_2$ symmetry (the RPa basis with $\ls 6=\ls 7 =0$); the explicitly CP symmetric model (the RPa basis with all parameters of the potential real), etc.

We find useful for our goals to analyze the model with known vacuum (the ground state of the potential) using the basis with $v_2 = 0$.
This basis is called {\it the Higgs (or Georgi) basis} \cite{Georgi}.
This basis is obtained from any given basis with   known v.e.v.'s  \eqref{genvac} by
transformation \eqref{reparam} with
\bear{c}
    \begin{pmatrix}{\Phi}_1\\ \Phi_2\end{pmatrix}\! =\!
    \hat{\cal F}_{HB}\!\begin{pmatrix}\phi_1\\ \phi_2\end{pmatrix},\;\; \hat{\cal F}_{HB}\!=\! \hat{\cal F}_{gen}(\theta\!=\!\beta, \tau\!=\!\rho\!-\!\xi).
\eear{reparamHB}
The phase factor $e^{\pm i\rho/2}$ represents the  remaining RPh freedom in the choice of the \HB\, that is,
independence of the physical picture from the choice of relative phase  $\phi_i$, the RPh phase.

{\it Vise versa,} any form of the potential can be obtained from the \HB form with the transformation,\lb
$\hat{\cal F}_{HB}^{-1}=\hat{\cal F}_{gen}(\theta=-\beta, \tau=\rho+\xi)$ with
 $\rho\to -\rho$, $\rho_0\to -\rho_0$.
We denote this basis as ``RPa basis $(\beta,\xi)/HB$''. Again, we do not fix in this definition the RPh phase $\rho$ and the irrelevant parameter $\rho_0$.

The potential obtained has the same form as \eqref{thdmpot}.
To distinguish its parameters in the \HB from a generic basis,
we use the capital letters $\Lambda$ for the quartic and $\mu$ for the quadratic parameters.
The extremum conditions  in the \HB are  simple,
$v^2\Lambda_1 = \mu_{11}^2$, $v^2\Lambda_6 = \mu_{12}^2$.
With these constraints the potential can be rewritten, up to a constant, in a more elegant form via the charged Higgs mass
$M_\pm^2$ \cite{GK07}
\bear{c}
\!\!V_{HB}\! =\! M_\pm^2\fft22\! +\! \dfrac{\Ls1}{2}\fftc11^2\!\!
\!+\!\dfrac{\Ls2}{2}\fft22^2\\[2mm]
+ \Ls3\fftc11\fft22+\Ls4\fft12\fft21\\[2mm]
        + \left[\dfrac{\Ls5}{2}\fft12^2+\Ls6\fftc11\fft12  \right.\\[2mm] \left.
 + \Ls7\fft22\fft12+\text{h.c.}\right].
\eear{HBmpot}

This fixing the RPa basis reduces the number of parameters needed to describe potential comparing to the original form \eqref{thdmpot}.
Instead of the four mass term parameters in its last line, we have two parameters,  the v.e.v. $v=246$~GeV and the mass of charged Higgs boson $M_\pm$.
The quartic part still contains 10 dimensionless parameters $\Ls{i}$ (including real and imaginary parts of $\Ls{5,6,7}$).
The residual RPh freedom is described with the irrelevant basis parameter $\rho$, the relative phase between the fields $\phi_i$.
The total number of relevant free parameters is 11, as mentioned above.

In the \HB, the decomposition \eqref{videf} simplifies to
\begin{equation}
\Phi_1=\left(\begin{array}{c}G^+\\ \dfrac{v+\eta_1+iG^0}{\sqrt2}\end{array}\right),
\quad
\Phi_2=\left(\begin{array}{c}H^+\\ \dfrac{\eta_2+i\eta_3}{\sqrt2}\end{array}\right)\,.
\label{decomp}
\end{equation}

To arrive to the description in terms of physically observable fields, one should
start by substituting these expressions into the potential \eqref{HBmpot}.
Also, by choosing the unitarity gauge for the gauge fields, we omit the
Goldstone modes $G^a$ from now on.

As a result, the potential \eqref{HBmpot} is takes the form in which coefficients are expressed via parameters of \eqref{HBmpot}
(here and below, the usual convention of summation over repeated indices is adopted):
\bear{rl}
V = &  M^2_{\pm}\,H^+H^- + \dfrac{M_{ij}}{2}\,\eta_i\eta_j \\[2mm]
 &  +vT_i\,H^+H^-\eta_i +vT_{ijk}\,\eta_i\eta_j\eta_k  \\[2mm]
&+CH^+H^-H^+H^- \\[1mm]
   & +\dfrac{B_{ij}}{2}\, H^+H^-\eta_i\eta_j +
     Q_{ijkl}\eta_i\eta_j\eta_k\eta_l.
\eear{Vindex}

%%%%%%%%%%%%%%%%%%%%%%%%%%%%%%%%%%%%%%
\section{Quadratic terms of potential (\ref{Vindex}). First group of observables}\label{ch.R}
%%%%%%%%%%%%%%%%%%%%%%%%%%%%%%%%%%%%%%%%%

In eq.~\eqref{Vindex}, the coefficients $M_{ij}$ form the neutral scalar {\it mass matrix} (here $N = M_\pm^2/v^2+\Ls4$):
\bear{l}
    M_{ij}\!=\! v^2\!
    \begin{pmatrix}
         \Ls1     & Re\,\Ls6             &-Im\,\Ls6          \\
         Re\,\Ls6 & \fr{N+Re\Ls5}{2}   &-Im\,\Ls5/2        \\
        -Im\,\Ls6 &-Im\,\Ls5/2           & \fr{N-Re\Ls5}{2}
    \end{pmatrix}\,.
\eear{massmatr}

The physical neutral Higgs states $h_a$ are such superpositions of fields $\eta_i$
that diagonalize this mass matrix:
\bea
&h_a=R_a^i\eta_i\,,\quad \eta_i = R^a_ih_a\,;&
\label{rotmatr}\\
&M_{ij}\eta_i\eta_j/2=\sum_a M_a^2h_a^2 /2,\quad M_{ij}=R^a_iR^a_jM_a^2.&
\label{Mdiag}
\eea

{\it The mixing matrix} $R^a_i$ is a real-valued orthogonal matrix determined
by the parameters of the mass matrix. It can be parameterized with three Euler
angles. One of them is responsible for rephasing transformation of fields, i.e. it is irrelevant.
The overall sign of this matrix is insignificant, we fix  $R^1_1>0$.

The trace of the mass matrix is invariant under orthogonal transformations
\eqref{rotmatr}. Therefore we obtain {\it a sum rule}:
\begin{equation}
v^2\left(\Ls1 +\Ls4\right)=\sum_a M_a^2- M_\pm^2\,.
\label{Mextract2}
\end{equation}

One of the advantages of the \HB as compared to other RPa bases is the fact that elements of rotation matrix are directly related to the couplings \eqref{relcoupldef}, \eqref{relcoupb}, which are, in principle, measurable:
\bear{c}
    \chi_{a}^V =R^a_1,\quad %\\[2mm]
     \chi_{a}^{H^+W^-}  \equiv \left(\chi_{a}^{H^-W^+}\right)^*= R^a_2+iR^a_3.\\[2mm]
      \eear{gcoupl}\vspace{0.2mm}
It can be seen easily after writing the kinetic term of Higgs Lagrangian with  definitions \eqref{decomp}, \eqref{rotmatr}.
The absolute values of the real quantities $\chi^V_a$ are directly measurable in the decays $h_a\to WW$ (or $W$-fusion process), etc.

The  phases of quantities $\chi_{a}^{H^+ W^-}$, \, i.e. the ratios $R^a_3/R^a_2$ cannot be fixed because of the rephasing freedom of potential in the \HB.
Their relative phases for different $h_a$ are determined unambiguously since they describe the physical quantity $\chi_{ab}^Z$ \eqref{Zhh}.
In particular, one can fix the rephasing phase (i.e. the RPh basis) by the condition $R^{2}_3=0$ (or some other condition).

The orthogonality of the mixing matrix means that its elements obey a set of relations:
\be
\sum\limits_i R^a_iR^b_i=\delta_{ab}
\,,\qquad \;\;\sum\limits_a R^a_iR^a_j=\delta_{ij}.
\label{orthm}
\ee
In a RPh basis with $R^{2}_3=0$, the relations \eqref{orthm} with $a=b$ and $i=j$ can be treated as equations determining some elements of mixing matrix via the others. Namely, we express all elements $R_i^a$ via couplings of different Higgs neutrals $h_a$ to gauge bosons $\chi_{a}^V$. The relative signs of some  elements of this matrix are given by basic expression of mixing matrix via Euler angles.  Including the phase rotation and restoring thus the rephasing freedom to allow for the phase $\rho$ in $\chi_{2}^{H^\pm W^\mp}$, we write general form of the mixing matrix:
\bear{c}
\!\!R_a^i\! =\!\!
    \begin{pmatrix}\!
        \chi_{1}^{V}\!&\!\chi_{2}^{V}\!&\!\chi_{3}^{V}\\[3mm]
        \dfrac{-\chi_{1}^{V}\chi_{2}^{V}}{\sqrt{1\!-\!(\chi_{2}^{V})^2}}&\sqrt{1\!\!-\!\!(\chi_{2}^{V})^2}&
        \dfrac{-\chi_{2}^{V}\chi_{3}^{V}}{\sqrt{1\!-\!(\chi_{2}^{V})^2}}\\
        \dfrac{\chi_{3}^{V}}{\sqrt{1\!-\!(\chi_{2}^{V})^2}}&0& \dfrac{-\chi_{1}^{V}}{\sqrt{1\!-\!(\chi_{2}^{V})^2}}\\
    \!\end{pmatrix} T,\\[3mm]
    T=\begin{pmatrix}
        1&0&0\\
        0&\cos\rho&\sin\rho\\
        0&-\sin\rho&\cos\rho\\
    \end{pmatrix}
\eear{Rexplicit}
with limitation, given by sum rule
\be
\sum_a(\chi_a^V)^2=1\,.
\label{SRV}
\ee

Finally, one can read \eqref{massmatr} as expressions of
some $\Lambda$'s via elements of the mass matrix and then, with the aid \eqref{Mdiag}, to express them via  the masses of Higgs bosons and their  couplings to gauge bosons (for $\LL4$ we   prefer to use the  sum rule  \eqref{Mextract2}):
\bear{lc}
v^2\LL1 =&\sum\limits_a (\chi^V_{a})^2M_a^2\,;\\[1mm]
 v^2\LL4   =   &
    \sum\limits_a M_a^2\!-\!M_{\pm}^2\!-\!v^2\LL1; \\[1mm]
 v^2\LL5^*
 =&\sum\limits_a (\chi_{a}^{H^+ W^-})^2M_a^2;\\[1mm]
 v^2\LL6^* =&\sum\limits_a \chi^V_{a}\chi_{a}^{H^+ W^-}M_a^2.
\eear{Mextract}
{ The observables entering into this equation form a first subset of the basic set of observables.} { Couplings $\chi_a^{H^+W^-}$ are expressed via $\chi_a^V$ by eq.~\eqref{Rexplicit}.}
The phase freedom in the definition of these couplings  is reproduced as a similar freedom in phases of $\LL5$, $\LL 6$.

%%%%%%%%%%%%%%%%%%%%%%%%%%%%%%%%%%%%%%%%%%%%%%%%%%%%%%%
\section{Other terms of potential}\label{ch.self}
%%%%%%%%%%%%%%%%%%%%%%%%%%%%%%%%%%%%%%%%%%%%%%%%%%%%%%%

The Higgs bosons masses and  couplings to the gauge bosons do not depend on $\LL2$, $\LL3$, $\LL7$. { In turn, these parameters are necessary to determine triple and quartic  Higgs bosons vertices. The measurements of some of these vertices are necessary to obtain complete set of observables. To construct second subset of observables, supplementing first subset to the basic set of variables (minimal and comprehensive),} we consider triple and quartic  interactions with the charged Higgs boson.
In each case we start our analysis in terms of fields $\eta_i$ and then pass to the physical Higgs fields $h_a$ with the aid of the mixing matrix \eqref{rotmatr}.

%%%%%%%%%%%%%%
{ \subsection{Cubic terms of potential (\ref{Vindex})}\label{cubic}
%%%%%%%%%%%%%%%%%

Each triple Higgs vertex depends on $\LL3$,  $Re\LL7$, $Im\LL7$, in addition to the  parameters of  the first subset.  }

\bu { We use for basic set the {\bf vertices $\pmb{H^+H^-h_a}$.}}
The simplest part of the cubic terms in \eqref{Vindex} describes interaction of neutral and charged scalars:
\begin{equation}
vT_i\,H^+H^-\eta_i,\quad\mbox{with}\quad T_i = \left(\LL3,Re\LL7,-Im\LL7\right)_i\,.
\label{defA}
\end{equation}
After transformation to physical states $\eta_i = R^a_ih_a$, we obtain the corresponding couplings:
$$
g(H^+H^-h_a) = vR^a_iT_i.
$$
This expression is easy to solve for $T_i$ by  inverting the rotation matrix.
In this way, we express  three parameters $\LL3$, $Re\LL7,\,Im\LL7 $
via the  measurable couplings between the neutral and charged scalars (cf. notation \eqref{relcouplc}):
\bear{c}
 \LL3= (2M_\pm^2/v^2)\sum\limits_a \chi_{a}^V\chi_a^\pm;\\[2mm]
 \LL7^*=
(2M_\pm^2/v^2)\sum\limits_a \chi_{a}^{H^-W^+}\chi_a^\pm\,.
\eear{triplecharged}

 \bu {\bf Vertices $\pmb{h_ah_bh_c}$} arise from the
part of  potential \eqref{Vindex}, cubic in neutral scalars:
\bear{c}
vT_{ijk}\eta_i\eta_j\eta_k = \dfrac{v}{2}\left[
\eta_1^3\,\Lambda_1+ \eta_1\eta_2^2\,(\Ls3\!+\!\Ls4\!+\!Re\Ls5)\right.\\[2mm]
+\eta_1\eta_3^2\,(\Ls3\!+\!\Ls4\!-\!Re\Ls5)-2 \eta_1\eta_2\eta_3\,Im\Lambda_5\\[2mm]
+3\eta_1^2(\eta_2\,Re\Lambda_6-\eta_3\,Im\lambda_6)\\[2mm]
\left.+\!(\eta_2^3+\eta_2\eta_3^2)\,Re\Lambda_7-(\eta_2^2\eta_3+\eta_3^3)\,Im\Lambda_7 \right]\,.
\eear{3neutral}
This equation contains parameters which are already familiar from quadratic terms and from eq.~\eqref{triplecharged}.
Passing as usual to the physical fields $h_a$, we transform this term to a form which exposes the triple neutral Higgs interactions $h_ah_bh_c$.

{ In the important particular case $b=c=a$, the
} using the orthogonality relations \eqref{orthm} gives \label{triple1}\bear{c}
g(h_ah_ah_a)=\\[2mm] v\left[\LL 1 (R_1^a)^3 +(\LL 3+\LL4)R_1^a\left(1-(R_1^a)^2\right)\right.\\[2mm] +Re\LL5 R_1^a\left((R_2^a)^2)-(R_3^a)^2\right) -2Im \LL5 R_1^aR_2^aR_3^a\\[2mm]
+3(R_1^a)^2(Re\LL6 R_2^a -Im \LL6 R_3^a)\\[2mm]+ \left.\left(1-(R_1^a)^2\right)(Re\LL7 R_2^a -Im \LL7 R_3^a)\right]\,.
\eear{triple1-a}

%%%%%%%%%%%%%%%%%%%%%%%
\subsection{Quartic terms of potential (\ref{Vindex})} \label{quartic}
%%%%%%%%%%%%%%%%%%%%

The parameter ${\LL2}$ can only be extracted from quartic couplings.
{ Each quartic Higgs vertex depends on  parameter $\LL2$
in addition to parameters determined from mass terms and triple Higgs couplings.

\bu  We  use for basic set {\bf the vertex $\pmb{H^+H^-H^+H^-}$}.}
In \HB, the charged Higgs bosons arise only from $\Phi_2$ \eqref{decomp}.
Therefore, the $H^+H^-H^+H^-$ vertex enters Lagrangian in a very simple form $\fr{\LL2}{2} H^+H^-H^+H^-$,
and its observation offers the simplest way to measure $\LL2$:
\begin{equation}
\LL2=   2g(H^+H^-H^+H^-)\,.
\label{l2_4ch}
\end{equation}

\bu {\bf The quartic scalar vertices involving neutrals $\pmb{h_a}$} are given by the sum
$\dfrac{B_{ij}}{2}\, H^+H^-\eta_i\eta_j+Q_{ijkl}\,\eta_i\eta_j\eta_k\eta_l $ in eq.~\eqref{Vindex}.
The coefficients here include the parameter $\LL2$, for example
\bear{c}
B_{ij} = \begin{pmatrix}\LL3  &Re\LL7 & -Im\LL7\\
                          Re\LL7& \LL2  &  0     \\
                         -Im\LL7& 0     &  \LL2  \end{pmatrix}
\eear{quartsun}
Experimental measurement of these vertices gives a cross check for the value of $\LL2$
obtained from the $H^+H^-H^+H^-$ interaction. The corresponding algebraic expressions
are straightforward but cumbersome.

\subsection{Other possible  choices}

{ The second subset of observables can be constructed  with other  triple  and  quartic couplings.}

The using of processes involving charged Higgses  looks preferable for two reasons. First, with charged Higgses, this procedure requires the fewest calculations, improving accuracy and reducing uncertainties. Second, the amplitudes of the processes $\epe\to H^+H^-h_a$, $\ggam\to H^+H^-h_a$,\lb $\epe\to H^+H^-H^+H^-$, $\ggam\to H^+H^-H^+H^-$
at ILC/CLIC \cite{HashemiAhmed} are  directly proportional to the corresponding couplings, without
any non-relevant diagrams interfering.

%%%%%%%%%%%%%%%%%%%%%%%%%%%%%%%%
\section{Yukawa interaction}\label{ch.Yuk}
%%%%%%%%%%%%%%%%%%%%%%%%%%%%%%%%%%%

The 2HDM can admit different forms of the Yukawa sector.
The interaction of a given right-handed down type fermion $f$ to neutral components of the Higgs fields can be written,
in the starting notation, as
$\Delta_f L_Y=\bar{f}_L\left(g_1\Phi_{1}^{0}+g_2\Phi_{2}^{0}\right)f_R+h.c.$,
where $\phi_{i}^{0}$ stand for the neutral component of $\phi_i$.
A simple RPa transformation $g_1\Phi_{1}^{0}+g_2\Phi_{2}^{0}=g_f\bar{\phi}_{1f}^{0}$
brings us to the  {\it f-selective} RPa basis,
which is adapted for this very fermion and
in which the mass-generating Yukawa interaction is written as
\be
\Delta_f L_Y= g_f \bar{f}_L\bar{\phi}_{1f}^{0} f_R+h.c.\label{Yukbas}
\ee
For a generic Yukawa sectors, these RPa bases are different for each fermion $f$. (For up-type quarks, the same expressions with $\Phi_i^*$ are assumed.)

In this  f-selective RPa basis,
the v.e.v.'s of Higgs fields have form \eqref{genvac} with parameters $\beta_f$, $\xi_f$, determined independently.
This basis is obtained from the Higgs basis form by a transformation inverse to \eqref{reparamHB},
i.e. that is $(\beta_f,\xi_f)/HB$ RPa basis.

Fixing $\rho_0=\rho/2$, we  express the field $\bar{\phi}_{1f}$ via the \HB fields $\Phi_i$
by equation $\bar{\phi}_{1f}=\cos\beta_f \Phi_1 -\sin\beta_f e^{i\xi_f}\Phi_2$.
In this basis, the v.e.v. of field $\bar{\phi}_{1f}$ is $v\cos\beta_f$,
which allows us to compactly write the corresponding couplings;
for example, $g_\phi^t=g_{SM}^t/\cos\beta_t$.
Now, decomposition of neutral components of $\Phi_i$ \eqref{decomp}
gives the interaction of these neutral component to $f$ in the form
\be
\!\!\Delta_f L_Y\!=\!g_f^{SM}
\fr{\bar{f}_ L\!\left[\cos\beta_f \eta_1\!\!+\!\!\sin\beta_f e^{i\xi_f}(\eta_2\!+\!\!i\eta_3)\right]\!f_ R\! +\!h.c.}{\cos\beta_f}\!\label{Yukbas1}
\ee
After that, the substitution of the rotation matrix \eqref{rotmatr}
gives couplings of down $f$-quark  to all neutral Higgses in the form
\be
\chi^f_a=\chi^V_a+\tan\beta_f e^{i\xi_f}\chi^{H^+W^-}_a\,.
\label{Yukcoupl}
\ee
In the CP conserved case these equations can be easily transformed to well known forms.

(For the up quark $f$ in the similar way one should write $\phi_i^*$ instead of $\phi_i$, it results in to the change $\xi\to -\xi$ in the final equations.)

It is known that a generic 2HDM Yukawa sector leads also to the flavor-changing neutral currents (FCNC). They arise due to mismatch of the f-selective bases of the quarks of the same charge.
If needed, these FCNC couplings can also be expressed, in a similar fashion, via the mismatch angles.

\bu It is instructive to illustrate this discussion with some discussed Yukawa sectors.

In 2HDM-I (Yukawa Model I), the f-preferable bases coincide for all fermions, $\beta_t=\beta_b$, $\xi_t=\xi_b$,
\bes\label{Yuk}\be
\chi^{u}_a=\chi^{d}_a=\chi^{\ell}_a\,.\label{YukI}
\ee
In 2HDM-II (Yukawa Model II), one such basis describes all up-quarks, and
an orthogonal basis describes all down quarks, $\beta_b=\pi/2-\beta_t\equiv\beta$, $\xi_b=\xi$, $\xi_t=0$, etc.
It leads to useful relations among Yukawa couplings for different fermions \cite{GK05}
\be
(\chi^u_{a} +\chi^d_{a})\chi^V_{a}=1+\chi^u_{a}\chi^d_{a}\,.
\label{YukII}
\ee\ees
In the aligned 2HDM \cite{aligned2HDM}, one has a similar picture --- one f-selective basis for all up quarks,
and another basis for all down quarks, --- but these two bases are not assumed to be orthogonal, { so that one can try to construct equations like \eqref{YukII} having more complex form.}

%%%%%%%%%%%%%%%%
\section{Some applications and by-products} \label{chappl}
%%%%%%%%%%%%%%%%

{  Our results allow to obtain a number of useful equations and to convert some known facts to a form more convenient for data analysis.

%%%%%%%%%%%%%%%%%%%%%%%
\bu {\bf Positivity constraints, etc.}
%%%%%%%%%%%%%%%%%%%%%
The positivity constraints can be now rewritten  via the measurable quantities from the basic set. For example, well known equation  $\sqrt{\LL1 \LL2}+\LL3>0$ is read now as
 \bear{c}
\sqrt{2g(H^+H^-H^+H^-)\sum\limits_a(\chi_a^V)^2M_a^2v^2}\,+\\+2M_\pm^2\sum\limits_a\chi_a^V\chi_a^\pm>0\,.
\eear{poscons}
Similar equations can be written for other positivity constraints and for the  perturbativity and unitarity  constraints, etc.\\

}

{ %%%%%%%%%%%%%%%%%%%%%%%%%%%%%
{\bf II. Coupling $\pmb{Zh_ah_b}$.}
%%%%%%%%%%%%%%%%%%%%%%%%%%%%%
The direct substitution of the rotation matrix into the kinetic term of Lagrangian gives    coupling $Zh_ah_b$ in the form}\black \\  \cl{$ \chi_{ab}^{Z} \equiv \dfrac{g(Z h_a h_b)}{M_Z/v} = R^a_2R^b_3-R^b_2R^a_3$.} This equation can be rewritten via couplings $\chi_{a}^{H^\mp W^\pm}$ and then -- with the aid of eq.~\eqref{Rexplicit} -- in the very simple form:
\bear{c}
 \chi_{ab}^{Z} =
  Im\left(\chi_{a}^{H^\mp W^\pm}\chi_{b}^{H^\pm W^\mp}\right)%\\[2mm]
   \equiv -\vep_{abc}\chi^V_{c}.
\eear{Zhh}
{ Here such equation, known for CP conserving case, is  spread for the most general case.}

%%%%%%%%%%%%%%%%%%%%%%
\bu {\bf III. Signature for CP conservation in 2HDM.}
%%%%%%%%%%%%%%%%%%%%%
In general, the neutrals $h_a$ have no definite CP parity.  The condition for CP conservation in the model { is written as a pair of almost obvious identities
\bes\label{CPchi}
\bea
{\black\prod\limits_a \chi^V_a=0\,;} \qquad
\prod\limits_a\chi_a^\pm=0\,.\label{CPchiV}
\eea
From Eq-s~\eqref{Mextract}, \eqref{triplecharged}  one can see  that these identities allow to have  potential in \HB with all real coefficients. It means that identities \eqref{CPchiV} form {\it necessary and sufficient conditions for CP conservation in 2HDM}. { In this case number of independent basic observables is reduced by 2 -- from 11 to 9, as it is known for the general explicitly CP-conserving case of 2HDM.}

The using of eqs.~\eqref{gcoupl}, \eqref{Rexplicit} and \eqref{Yukcoupl} shows that  the first condition \eqref{CPchiV} ensures the validity of the necessary condition of the CP conservation for each fermion
\be
\left|\prod\limits_a \chi^f_a\right| =\prod\limits_a \left|\chi^f_a\right|\,.\label{CPchif}
\ee
\ees

Let us note  that even weak violation of CP for coupling $h_a$ to gauge boson is compatible with strong enough violation of CP in the $\bar{f}h_af$ interaction  (if corresponding $\tan\beta_f$ (or $\cot\beta_f$) is large).}\\

%%%%%%%%%%%%%%%%%%%%%%%%%%%%%%%
{\bf IV. Sum rules for Higgs couplings.}
%%%%%%%%%%%%%%%%%%%%%%%%
It is useful, in addition to \eqref{Rexplicit}, to rewrite equations \eqref{orthm}  at $a=b$ and at $i=j$  and eq.~\eqref{Yukcoupl} in the form of sum rules,
allowing for generalizations to some other forms of the Higgs sector \cite{Gin14}. That are relation \eqref{SRV} and
\bear{c}
(a)\; \sum\limits_a (\chi_a^f)^2=1\,,\;\;%\\[2mm]
(b)\, |\chi_{a}^V|^2+| \chi_{a}^{H^+W^-}|^2=1
 \,.
\eear{SR}\black
Relations \eqref{SRV}  and (a) are the sum rules for couplings of different neutral scalars to gauge bosons and fermions. They are well known for the CP conserving case and
for some definite forms of the Yukawa sector (see e.g. \cite{gunion-haber-wudka,Celis}).
The main new point here is the statement about validity of all these sum rules beyond CP conservation and for an arbitrary form of the Yukawa sector.

The relations (b) for $a=1, 2, 3$ represent a new set of sum rules, which is useful for assessing the physics potential of the forthcoming experiments.\\

%%%%%%%%%%%%%%%%%%%%%%%%%
{\bf V. Possible strong interaction in the Higgs sector.}
%%%%%%%%%%%%%%
The fact that free parameters of the potential naturally fall into three very distinct categories, offers a new opportunity which was absent in the SM. Before the Higgs discovery, the large coupling constant $\lambda$ was, in principle, possible within SM. In this case, the Higgs boson would be very heavy and wide, and it could not be seen as separate particle. Instead, its dynamics would be governed by the strong interaction in the Higgs sector, which would manifest itself in the form of resonances in the $W_LW_L$, $W_LZ_L$, $Z_LZ_L$ scattering in the 1-2 TeV energy range. In the SM this opportunity is closed by the discovery of the Higgs boson with $M\approx 125$~GeV.

Our analysis shows that, within 2HDM, the reasonably low values of all Higgs masses are well compatible
with large $\LL3$, $|\LL7|$, $\LL2$, i.e. with the strong interaction in the Higgs sector.
A signal of this feature can be observed in the multi-Higgs final states or (for $\LL3$, $|\LL7|$) in anomalously large two-photon width of some neutral Higgs boson.
Moreover, this strong interaction can coexist even with moderate values of triple Higgs couplings  as it could be driven
exclusively by the large value of a single parameter $\LL2$.\\

%%%%%%%%%%%%%%%%%%%%%%%%%%%%%%%%%%%%%%%%%%%%%%%%%%%%%%%%%%%
{\bf VI. Triple Higgs vertex.}
%%%%%%%%%%%%%%%%%%%%%%%%%%%%%%%%%%%%%%%%%%%%%%%%%%%%%%%%%%
{ To express triple Higgs vertex \eqref{triple1-a} in terms of observables we use \eqref{Mextract} and \eqref{triplecharged}. We obtain}\black
\bear{c}
g(h_ah_ah_a)=(M_a^2/v)\;\chi_{aaa};\\
\chi_{aaa}\!=\!
\chi^V_a\!\left\{1\!+\!\left(1\!-\!(\chi^V_a)^2\right)\!\!\left[1
+\sum\limits_b 2\fr{M_b^2}{M_1^2}(\chi^V_b)^2\right]\right.\!+\!\\[2mm]
\left.+\!\left(1-(\chi^V_a)^2\right)\fr{2M_\pm^2}{M_a^2}N\right\}\\[3mm]N\!=\!
\sum\limits_b\chi^V_b\chi^\pm_b\!-\!1\!+\!Re\left(\sum\limits_b
\chi^{H^+W^-}_b\chi^\pm_b\fr{\chi^{H^+W^-}_a}{\chi^V_a}\right).
\eear{triple1-c}

For $a=1$, the { first factor $M_1^2/v$} is similar to what we would get in the SM  when describing the triple Higgs vertex.  This equation is used in \cite{Gintriple} for estimating opportunity to observe sizable violation of this vertex from its SM value { in the case of realization of SM-like scenario.\\

%%%%%%%%%%%%%%%%%%%%%%%%%%%%%%%%%%%%%%%%%%%%%%%%%%%%%%%%%%%
{\bf VII. Notes about radiative corrections.}
%%%%%%%%%%%%%%%%%%%%%%%%%%%%%%%%%%%%%%%%%%%%%%%%%%%%%%%%%%

The standard calculation of the radiative corrections (RC) in the model is based on the parameters of Lagrangian which are RPa dependent.  This RPa ambiguity can be removed, for example, by using the renormalization procedure fixing parameters of the basic set. In the modern approach the calculation of any physical effect should be supplemented by calculation of renormalized values of masses and other parameters of basic set which should be taken into account in the data analysis\fn{ For example, in some particular variant of MSSM the value of triple Higgs coupling with RC looks essentially different from its tree form in the SM \cite{hhhRC}. However, within the same approximation the using of the renormalized mass $M_1$
makes the result  close to the SM value  \cite{Boudhhh}.}.

Besides, with anticipated low accuracy in the determination of parameters of Lagrangian from future data, the calculation of RC could  help in the understanding of  the  model if only  $RC\gtrsim 10$\%. These big RC can appear  in the case of strong interaction in the Higgs sector (either direct or via $t$-quarks).}

%%%%%%%%%%%%%%%%%%%%%%%%%%%%%%%%%%%%%%%%%%%%%%%%%%%%%%%%
\section{Discussion} \label{ch.disc}
%%%%%%%%%%%%%%%%%%%%%%%%%%%%%%%%%%%%%%%%%%%%%%%%%%%%%%%%

{ $\lozenge$  We have found the  minimal complete set of measurable quantities (named {\it observables}) which determines  all parameters of the 2HDM  Lagrangian -- {\it the basic set of observables}. This set is naturally subdivided into two subsets.

The first subset   contains   masses of all Higgs bosons $M_{1,2,3}$, $M_\pm$, vacuum expectation value of Higgs field $v=246$~GeV and the couplings $\chi^V_a$ of any   two  (of   three)  chosen  neutrals to the gauge bosons. Here we assume that $h_1$ is  the  discovered Higgs boson with $M_1\approx 125$~GeV. The final equations also contain  the  couplings $\chi_a^{H^+W^-}$, expressed via $\chi_a^V$ with the aid of eq.~\eqref{Rexplicit}. This subset determines  explicitly all quadratic (mass) terms  of potential \eqref{HBmpot}.

The  coefficients $\LL1,\,\LL4,\, \LL5,\,\LL6$ of Lagrangian are expressed simply via observables of the first subset \eqref{Mextract}.

$\lozenge$ The triple and quartic Higgs vertices of the potential \eqref{HBmpot}  can be determined completely  only if one supplements  the parameters of the first subset by an additional information. In turn, to form the second subset, one  need to use triple and quartic Higgs self-interactions. For this goal we use  three triple couplings $H^+H^-h_a$ (quantities $\chi^\pm_a$) and one quartic coupling $g(H^+H^-H^+H^-)$.

The parameters of the first subset plus three couplings $\chi^\pm_a$ determine all triple Higgs couplings. The coefficients $\LL3,\,\LL7$ of Lagrangian are expressed simply via these three couplings and  observables of the first subset.

The description of quartic interactions of Higgs particles demands to add one more observable $g(H^+H^-H^+H^-)=\LL2 /2$.}

\bu { The obtained equations for parameters of Lagrangian in \HB} contain one irrelevant parameter: the RPh phase $\rho$ related to  a rephasing freedom in the \HB. In order to switch to another RPa basis, which could be more useful for some special reasons, one should use two  parameters $\tan\beta$ and $\xi$, which are determined by the RPa basis choice. Once these parameters are determined from  problem-specific conditions, the transition to this RPa basis is performed with the aid of the back rotation $\hat{\cal F}_{HB}^{-1}$ \eqref{reparamHB}, as it was done in Sect.~\ref{ch.Yuk}. The final equations for parameters $\ls i$, $m^2_{ij}$ are constructed from measurable quantities discussed above and  RPa basis-choice parameters\fn{ Within Higgs sector all these parameters are non-physical since they describe only RPa freedom.  The choice of f-selective RPa basis, in which the mass-generating Yukawa interaction of definite fermion $f$ is written in the form \eqref{Yukbas}, fixes value $\beta=\beta_f$. The parameter $\beta$ accepts a meaning if it is the same for different fermions
(or values $\beta_f$ for different fermions are simply related to each other) -- as in Models I or II.} $\beta$, $\rho$, $\xi$.

\bu The observables of the basic set are  measurable quantities,
independent of each other. The models with arbitrary values of these observable parameters can in principle be realized, provided that the positivity constraints like \eqref{poscons} are satisfied and  the couplings $\chi^V_a$ are not too large, in order not to violate    the   sum rule \eqref{SRV}. In some special variants of 2HDM,  additional relations between these parameters may appear (for example, in the CP conserving case $\chi_3^V=\chi_3^\pm =0$).

{ Our results open  the  door for the study of  Higgs models in terms of measurable quantities alone. It allows to remove from the data analysis the widely spread intermediate stages  with complex, often model-dependent, analysis of  coefficients of Lagrangian.
}

{ \bu The principal possibility to determine all parameters of 2HDM from the (future) data meet  strong practical limitations (which can be hidden in other approaches). In the best case, it  looks  the problem for  a  very long time.

Indeed, the  modern data on  the  Higgs boson couplings,
the  analysis of many particular models (see e.g. \cite{genref}),  and the   using of sum rules \eqref{SRV}.  \eqref{SR} allow  us  to conclude that the discovery of new Higgs bosons $h_{2,3}$, $H^\pm$ is  a  difficult problem for LHC and $e^+e^-$ colliders  \cite{Gin15-1}.

If these $h_{2,3}$ are discovered,  the  inaccuracies
in the measuring of their masses and couplings are expected to be  not small.}

{ The measuring of triple and quartic interactions of Higgs bosons looks more difficult problem. So that it is natural to expect that these measurements will be made later and with bigger inaccuracy.

The general limitations for model, similar to the positivity constraint  \eqref{poscons}, contain parameters of the first and second subsets simultaneously.   Thus, there are   few chances that such restrictions  can be verified   in  the  near future.\\
  }

This work was supported  by grants RFBR 15-02-05868, NSh-3003.2014.2 and  NCN OPUS 2012/05/B/ST2/03306 (2012-2016).
We are thankful to  I.P.~Ivanov and M. Krawczyk for discussions.

%%%%%%%%%%%%%%%%%%%%%%%%%%%%%%%%%%%%%%%%%%%%%%%%%%%%%%%%%%%%%%%%%%%%%%%%%%%%%%%%%%%%%%%%%%

\end{document}